\documentclass[mathleft
]{myan}
\usepackage{graphicx}
\usepackage{times}
\begin{document}

\Pagespan{1}{3}
\Yearpublication{2008}%
\Yearsubmission{2007}%
\Month{3}%
\Volume{329}%
\Issue{3}%

\title{The Palomar-Quest Digital Synoptic Sky Survey
}

\author{S.G. Djorgovski\inst{1},\fnmsep\thanks{Corresponding author:
  Djorgovski, \email{george@astro.caltech.edu}\newline}
C. Baltay\inst{2},
A.A. Mahabal\inst{1},
A.J. Drake\inst{3},
R. Williams\inst{3},
D. Rabinowitz\inst{2},
M.J. Graham\inst{3},
C. Donalek\inst{1},
E. Glikman\inst{1},
A. Bauer\inst{2},
R. Scalzo\inst{2},
N. Ellman\inst{2},
\and J. Jerke\inst{2}
}
\titlerunning{Palomar-Quest Survey}
\authorrunning{S.G. Djorgovski et al.}
\institute{
Astronomy, MS 105-24, Caltech, Pasadena, CA 91125, USA
\and 
Physics Dept., Yale University, New Haven, CT 06520, USA
\and 
Center for Advanced Computing Research, MS 158-79, Caltech, Pasadena, CA 91125, USA
}

\received{01 Sep 2007}
\accepted{25 Dec 2007}


\abstract{
We describe briefly the Palomar-Quest (PQ) digital synoptic sky survey, including its parameters, data processing, status, and plans.  Exploration of the time domain is now the central scientific and technological focus of the survey.  To this end, we have developed a real-time pipeline for detection of transient sources.  We describe some of the early results, and lessons learned which may be useful for other, similar projects, and time-domain astronomy in general.  Finally, we discuss some issues and challenges posed by the real-time analysis and scientific exploitation of massive data streams from modern synoptic sky surveys.
}

\maketitle

\section{A Brief Description of the PO Survey}
The Palomar-Quest (PQ) digital synoptic sky survey is a collaborative project between groups at Yale University and Caltech (Co-PIs: C. Baltay and S.G. Djorgovski), with an extended network of collaborations with other groups world - wide, including Indiana U. (M. Gebhard et al.), NCSA (R. Brunner et al.), LBNL Nearby SN Factory (NSNF; S. Perlmutter et al.), INAOE (Puebla, Mexico; L. Carrasco, O. Lo\-pez-Cruz et al.), EPFL (Switzerland; G. Meylan et al.), and Caltech/JPL (M. Brown et al.).  The data are obtained at the Palomar ObservatoryÕs Samuel Oschin telescope (the 48-inch Schmidt) using the QUEST-2 112-CCD, 161 Mpix ca\-mera (Baltay et al. 2007).  Approx. 45\% of the telescope time is used for the PQ survey.  The survey started in the late summer of 2003, and will finish in the late 2008.

In the first phase of the survey, data were obtained in the drift scan mode in $4.6^\circ$ wide strips of a constant Dec, in the range $-25^\circ < \delta < +25^\circ$, excluding the Galactic plane.  The total area coverage is $\sim 15,000~deg^2$, with multiple passes, ranging from a few to about 25, and typically 5 -- 10 times, with time baselines ranging from hours to years.  There are some thin-strip gaps in the coverage, due to a combination of inter-CCD gaps, bad CCDs, and a suboptimal dithering strategy.  Typical area coverage rate is up to $\sim 500~deg^2$/night in 4 filters.  The raw data rate is on average $\sim 70$GB per clear night.  To date, about 25 TB of usable data have been collected in the drift scan mode. 

Data were obtained with two filter sets, Johnson $UBRI$ and Gunn/SDSS $rizz$, recently changed to $griz$.  Effective exposures are $\sim 150$ sec / cos $\delta$ per pass.  Typical estimated limiting magnitudes are 
$r_{lim} \approx 21.5$,
$i_{lim} \approx 20.5$,
$z_{lim} \approx 19.5$,
$R_{lim} \approx 22$, and 
$I_{lim} \approx 21$ mag, depending on the seeing, lunar phase, etc.
Coadding of $\sim 8$ passes reaches the depth of SDSS in the redder bands.
Photometric calibrations are done independently at Yale and Caltech, mainly using the overlap region with SDSS.

In the second phase of the survey, which started in the spring of 2007, data are obtained in the traditional point-and-track mode, in a single, wide-band red filter (RG610), with $\sim 10$\% of the time in the drift scan mode.  The coverage and the cadence are optimized for the nearby supernova search, in collaboration with the LBNL NSNF group, and a search for dwarf planets, in collaboration with M. Brown.

Data are processed with several different pipelines, optimized for different scientific goals.  This includes the Yale pipeline (Andrews et al. 2007), which does the PSF fitting and was designed for a search for gravitationally lensed quasars; the Caltech data cleaning pipeline, used to remove numerous instrumental artifacts present in the data; the Caltech real-time pipeline, used for real-time detections of transient events, as described below; the LBNL NSNF pipeline, based on image subtraction and designed for detection of nearby SNe; and a pipeline for an optimal coadding of images and detection of sources in them, now developed at Caltech.  Images and resulting catalogs are stored in multiple locations, using a variety of databases.

PQ is the first major digital sky survey fully designed and implemented in the Virtual Observatory (VO) era, and it uses VO standards and protocols throughout.  Public data releases will be also done through VO-type interfaces.  The first public data release is imminent, pending the completion of various data quality control and assessment tests.

The survey is feeding multiple scientific goals and pro\-jects.  The initial motivation was a search for $> 10^5$ QSOs, using colors and variability, in order to discover $> 100$ strong gravitational lenses, and use them to constrain cosmology and/or history of mass assembly.  Another project was a search for high-$z$ QSOs, to be used as probes of reionization and early structure formation.  Both of them are now finally starting to yield results; the progress was slow due to numerous problems with the data, all of which have been solved, and will be documented in detail elsewhere.  Our principal scientific focus now is exploration of time domain, as described below.

Our main public outreach effort to date has been the creation of the Griffith Observatory's ``Big Picture'', and the associated website, http://bigpicture.caltech.edu.  This exhibit will be seen by millions of visitors, serving multiple educational roles in the years to come.

\section{PQ Exploration of the Time Domain:\\
~~~~~Some Preliminary Results}
With a data set covering nearly 40\% of the entire sky, with multiple passes reaching $\sim 21$ mag each, and time baselines ranging from minutes (between different CCDs) to hours (repeated scans in the same night), days (within the same lunation), months, and years, and (using the cross-matches to DPOSS and SDSS catalogs) up to decades, PQ is in a unique position to explore time-variable sky in a systematic fashion.  For some early reports, see Graham et al. (2005), Mahabal et al. (2004, 2005), or Djorgovski et al. (2006).

One major effort is a search for nearby ($z \sim 0.1$) SNe Ia, to be used as the low-$z$ calibration of the Hubble diagram.  This project is led by the Yale group in collaboration with the LBNL NSNF.  To date, this effort has found a total of about 500 SNe, about a half of which were spectroscopically confirmed, and among them about 70 Type IaÕs with 10 or more spectra taken; as well as a plethora of other SNe (including some peculiar ones) and transients.  All are published in IAU Circulars, CBETs, and ATel's.  The work uses image subtraction technique, in order to remove the well-detected light host galaxies.  The Caltech real-time pipeline is now also starting to detect SNe, using a search for transients in the catalog domain.  

We are now using the archives of our data to study systematically the variability of QSOs, and especially Blazars.  Some examples are shown in Fig. 1.  The main goal is to devise an algorithm based on colors and variability alone to define a purely optically selected sample of Blazars, and thus check on the selection effects in the traditional radio and x-ray approaches.  These sources may be the main contributors to the extragalactic $\gamma$-ray background, a subject of considerable interest with the upcoming launch of the GLAST mission.  They are also implicated as sources of TeV-scale (and presumably even more energetic) $\gamma$-rays and ultra-high energy cosmic rays (UHECR).  These cosmic accelerators can reach energies several orders of magnitude higher than any predictable terrestrial accelerators.  Their census and detailed studies are thus of a considerable and growing interest.

\begin{figure}
\includegraphics[width=70mm]{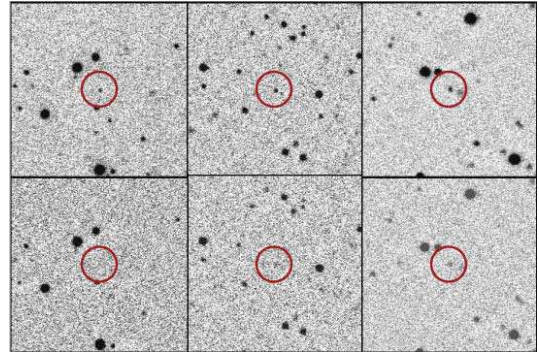}
\caption{
Examples of 3 known Blazars, as seen in the PQ data.  The top row shows them in a relatively high state, the bottom row in a relatively low state.
}
\label{label1}
\end{figure}

Our exploration of the archival PQ data has yielded a large number of transients, operationally defined as PSF-like sources detected in only one epoch, with no detectable apparent motion between different CCDs in a single pass.  Subsequent studies have revealed counterparts for some of them in deeper, coadded images.  We believe that many of them are probably asteroids caught near the stationary point (see below).  However, this has underscored the need to detect and follow transients in a real or near-real time, in order to determine their physical nature.

We have thus developed a real-time pipeline, which is now operational.  The pipeline does the standard removal of instrumental signatures, pushes the data through the Caltech cleaning pipeline, detects and measures sources, implements astrometry, compares the new catalogs to those from the previous passes, finds newly detected sources, implements a number of software filters to eliminate the residual instrumental artifacts, known asteroids or variables, moving objects (uncatalogued asteroids), produces cutout images and webpages for the candidate transients, and publishes them using the VOEvent protocols and on VOEN website,
http://voeventnet.caltech.edu/feeds/PQ\_OT.shtml

\begin{figure}
\includegraphics[width=65mm]{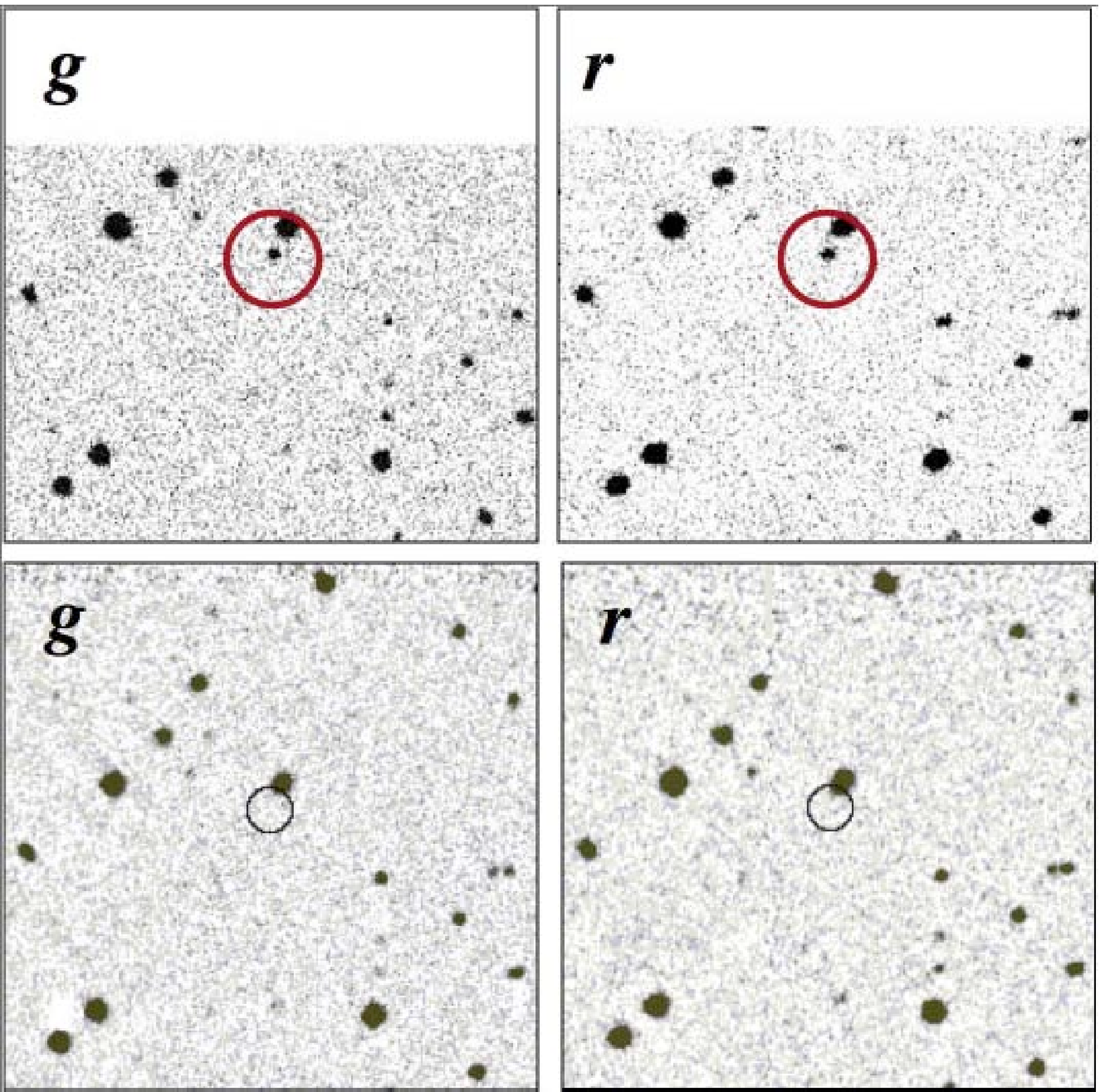}
\caption{
An example of a transient detected with our real-time pipeline, PQT 070519:143304+150707; see Drake et al., ATel 1083.  The top row are detection images in the $g$ and $r$ bands; the bottom row are the comparison baseline images.  The source faded slowly, but got redder rapidly; it may be a rare type of a SN.
}
\label{label2}
\end{figure}

We typically do a $\sim 4$-hour long scan, then re-scan the same area again, with the real-time pipeline running.  In a typical half-night scan, we may detect a couple of million sources, and about a thousand potential transients.  Removal of residual instrumental artifacts leaves a few hundred genuine detections, nearly all of which are asteroids; of them, typically only a half are among the previously catalogued; the rest are largely removed after the second scan.  
The number of VOEvents generated has been evolving as the filtering procedures were improved.  Over the past year or so, nearly 4800 events have been submitted, with an average rate of $\sim 200$ per night.  About 85\% of these were immediately classified as asteroids, and the majority of the remaining ones are as well.
Finally, there are only a few ($< 10$/night) apparently genuine astrophysical transients left.  Spectroscopic follow-up observations to date show that they are a mixture of SNe, AGN, probable flaring M dwarfs, and the rest are of as yet unknown nature.  Some are re-discovered on different nights.

\section{Some Lessons Learned}
Combining the current PQ experiences with the older work with DPOSS (see, e.g., Mahabal et al. 2005), we estimate that in a single-pass snapshot survey there are $\sim 10^{-2}$ astrophysical transients/$deg^2$ down to $\sim 20$ mag at high Galactic latitudes.  Many of them are known, highly variable types of objects, where the ``low state'' is below the detection of the baseline data, with variable stars of different kinds dominating on the short time scales ($\sim$ minutes to months), and AGN (mainly Blazars and OVVs) dominating on the longer time scales (years and longer).  Some are a variety of stellar explosions.  Some may be as-yet unknown types of objects and phenomena, but real-time spectroscopic and other follow-up is necessary in order to discover them.

We find that a principal contaminant for optical surveys are the slow-moving asteroids; there are $\sim 1 - 3$ of them per $deg^2$ down to $\sim 21$ mag, depending very much on the Ecliptic latitude; i.e., $> 100$ asteroids for each astrophysical transient.  A joint analysis for moving and variable objects is necessary, and any type of a synoptic sky survey data stream can feed both scientific domains simultaneously.  Improving the existing catalogs of asteroids is an urgent task.  At least two epochs are needed in order to eliminate previously unknown asteroids in any synoptic survey, and their baseline will define the effective time resolution of any transient search (we also note that at least 3 properly spaced epochs are needed to compute even a rough orbit).

The quality of the baseline or fiducial sky against which current observations are compared is a key issue.  It must be deep, clean, complete, and wavelength-matched.  Generating a standard, dynamically evolving, annotated, multi-wavelength baseline sky may be a good community (VO) project; we are developing a prototype from PQ and other publicly available panoramic imaging data sets.

Achieving a high completeness (a few real transients mis\-sed) and a low contamination (a few false alarms) is a huge challenge.  Interesting sources are discovered as outliers in some parameter space; problems with the data also generate outliers in some parameter space.  In a large data set, most unlikely things will happen, and most of them are bad.  Robust and reliable data cleaning is a key requirement.  This is hard to do in a cutting-edge software system.

Data systems (pipelines, archives, and analysis) and operational procedures for synoptic sky surveys are subject to a substantial tension between static and dynamic components, including both real-time and subsequent (non-time-critical) analysis and distribution, data ingestion, database updating and recomputing, etc.  This has implications both for survey strategies and system architecture design.

Another key challenge is an automated classification of events for prioritized follow-up, as discussed by Mahabal et al. and Bloom et al. elsewhere in this volume.  This will certainly require use of machine learning tools, as described by Vestrand et al. in this volume.

All of these challenges will grow much sharper, as the data volume and data flux increases dramatically in upcoming synoptic sky surveys.  We are now dealing with data streams of the order of 0.1 TB/night, and $\sim 10$ transients/nt.  On a time scale of $\sim 1 - 5$ years, this will increase to $\sim 1$ TB/night and $\sim 10^4$ transients/night (e.g., PanSTARRS), and on a time scale of $\sim 5 - 10$ years, this will increase to $\sim 20$ TB/night and $\sim 10^5 - 10^6$ transients/night (LSST).  Development and testing of software, methodologies, and operational and follow-up procedures is an urgent task, in which surveys such as PQ can play an important role.

\acknowledgements
We thank many collaborators who have made essential contributions to the survey, and the staff of Palomar Observatory for their tireless efforts during the survey operations.  This work was supported in part by the NSF grants AST-0407448, AST-0326524, and CNS-0540369, by the Ajax Foundation, and other private donors.  SGD acknowledges a stimulating atmosphere of the Aspen Center for Physics.  Finally, we thank the workshop organizers for an excellent and productive meeting.


\begin{thebibliography}{}
  \bibitem{} Andrews, P., et al.: 2007, PASP, in press (astro-ph/0703446)
  \bibitem{} Baltay, C., et al.: 2007, PASP, in press (astro-ph/0702590)
  \bibitem{} Djorgovski, S.G., et al.: 2006, in Proc. ICPR2006, eds. Y.Y. Tang et al., IEEE Press, p. 856 (astro-ph/0608638)
  \bibitem{} Graham, M., et al. (the PQ Survey Team): 2004: in Proc. ADASS XIII, eds. F. Ochsenbein et al., ASPCS 314, 14
  \bibitem{} Mahabal, A., et al.: 2004, in press (astro-ph/0408035)
  \bibitem{} Mahabal, A., et al. (the PQ Survey Team): 2005, in Proc. ADASS XIV,
eds. P. Shopbell et al., ASPCS 347, 604

\end{thebibliography}
\end{document}